%% file: main.tex
\documentclass[12pt, letterpaper]{article}

\usepackage{graphicx}
\usepackage{todonotes}
\usepackage{xurl}

\AtBeginDocument{%
  \providecommand\BibTeX{{%
    \normalfont B\kern-0.5em{\scshape i\kern-0.25em b}\kern-0.8em\TeX}}}

\begin{document}

\title{SoK: Not Quite Water Under the Bridge: Review of Cross-Chain Bridge Hacks}

\author{Sung-Shine Lee, Alexandr Murashkin, Martin Derka, Jan Gorzny}
%\author{Sung-Shine Lee}
%\email{martinet@quantstamp.com}
%\author{Alexandr Murashkin}
%\email{alex@quantstamp.com}
%\author{Martin Derka}
%\email{martin@quantstamp.com}
%\author{Jan Gorzny}
%\email{jan@quantstamp.com}
%\affiliation{%
%  \institution{Quantstamp, Inc.}
%  %\streetaddress{P.O. Box 1212}
%  %\city{Dublin}
%  %\state{Ohio}
%  \country{USA}
%  %\postcode{43017-6221}
%}

%\renewcommand{\shortauthors}{Lee et al.}


%\keywords{bridges}

\maketitle

\begin{abstract}
  The blockchain ecosystem has evolved into a multi-chain world with various blockchains vying for use.
  Although each blockchain may have its own native cryptocurrency or digital assets, there are use cases to transfer these assets between blockchains.
  Systems that bring these digital assets across blockchains are called \emph{bridges}, and have become important parts of the ecosystem. 
  The designs of bridges vary and range from quite primitive to extremely complex. 
  However, they typically consist of smart contracts holding and releasing digital assets, as well as nodes that help facilitate user interactions between chains. 
  In this paper, we first provide a high level breakdown of components in a bridge and the different processes for some bridge designs.
  In doing this, we identify risks associated with bridge components.
  Then we analyse past exploits in the blockchain ecosystem that specifically targeted bridges.
\end{abstract}

\sloppy
\input{sections/intro}
\input{sections/bridges}
\input{sections/custodian}
\input{sections/debt-issuer}
\input{sections/communicator}
\input{sections/interface}
\input{sections/related}
\input{sections/conclusion}
\input{sections/acknowledgement}

%\bibliographystyle{ACM-Reference-Format}
\bibliographystyle{unsrt}
\bibliography{references}

\end{document}

%% file: sections/intro.tex
\section{Introduction}\label{sec:intro}

In recent years, the blockchain ecosystem has evolved into a multi-chain world. 
Various blockchains, like the popular Bitcoin Network \cite{bitcoin} or Ethereum~\cite{ethereum}, are evolving simultaneously.
These blockchains often have their own native cryptographically-based digital asset or cryptocurrency, like Bitcoin or Ether.
Advanced blockchains like Ethereum support automatically executed pieces of code, so-called \emph{smart contracts}, which enable programs to be developed on these blockchains.
In turn, these programs often introduce additional digital assets, like non-fungible tokens (NFTs).

However, for various reasons, it is often desirable to move digital assets from one blockchain to another.
For example, a user may wish to move their Bitcoin onto Ethereum to deposit it into various \emph{Decentralized Finance} (DeFi) protocols which may allow the user to earn interest on their Bitcoin, akin to a savings account from a bank (see e.g., \cite{defi-intro}).
In another situation, a user may have an Ethereum-based NFT which can be used in simulated races executed by smart contracts.
However, the gas fees -- the cost of executing a transaction on Ethereum -- may be prohibitively large, and so the simulated race may be built on a so-called \emph{layer two} scaling solution built on top of Ethereum \cite{cryptoeprint:2019/360}.
Examples of these scaling solutions include \emph{rollups} (also known as \emph{commit-chains} \cite{cryptoeprint:2018/642}), \emph{Plasma} \cite{plasma}, or a side-chain (see e.g., \cite{sidechains}).
These scaling solutions have their security tied to Ethereum but are able to reduce the cost of gas fees, and are therefore more attractive for applications such as an NFT-based simulated race.

\begin{figure}
    \centering
    \includegraphics[scale=0.65]{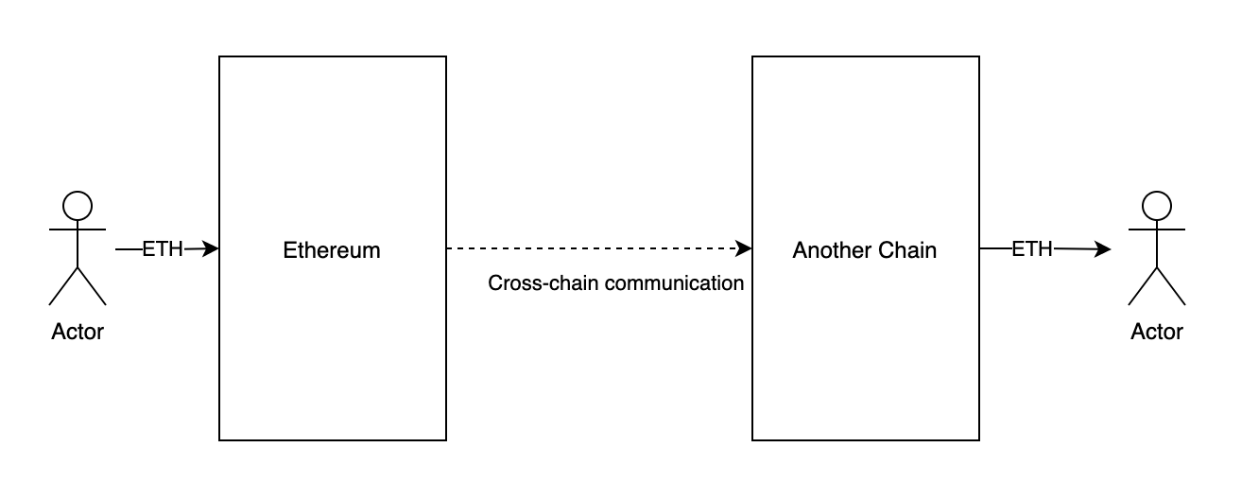}
    \caption{Cross-chain communication illustrated, from the Ethereum blockchain to another blockchain, ``Another Chain.'' In this case, the actor on the left wants to send Ether (ETH), the native digital asset for Ethereum, to another actor on the right, who uses Another Chain. The dotted line between the blockchains requires cross-chain communication, or a bridge.}
    \label{fig:cross-chain-communication}
\end{figure}

The ability to ``move'' a digital asset from one blockchain to another requires a protocol, which can be implemented with the support of smart contracts.
Such a protocol is required to ensure that an asset can only be used on one blockchain at a time, in order to prevent \emph{double spending}. 
Double spending is when one spends a cryptocurrency token twice \cite{DBLP:journals/access/IqbalM21}. 
In this case, the double spend would be one spend of the token on its original blockchain and one spend of the token on the blockchain it was moved to.
However, since a blockchain is a self-contained system, it is impossible to actually take a digital asset from one source blockchain and put it on a different destination blockchain. 
Instead, a representation of the original asset on the source blockchain must be created on the destination blockchain.
Thus such a protocol must involve some cross-chain communication (see Figure \ref{fig:cross-chain-communication}), and this communication protocol (and its implementation) is called a \emph{bridge}.

Bridges are complicated protocols and software projects. 
They can be as complicated as blockchains themselves (especially if they are a decentralized protocol), and may be required to be implemented in various languages (one for each blockchain the bridge interacts with), each with its own nuances.

Moreover, bridges need to unlock or mint digital assets on destination blockchains. 
Bridges are therefore responsible for distributing valuable digital assets, and as a result, have become targets for attackers.
In the last year, over \$1 billion USD worth of digital assets have been stolen, incorrectly minted, or locked in these systems \cite{wiredattacksdata}. 
The result is that some users are without their digital assets, confidence in cross-chain protocols is shaken, and protocols fail to operate as promised.

In order to prevent issues like this from happening again, a deep understanding of how bridges work is required.
In Section \ref{sec:bridges}, we review the general structure of a bridge.
Next, we review various bridge exploits that have occurred on the components of a bridge.
In Section \ref{sec:custodian}, we recount exploits that have or would have targeted parts of the bridges that hold user assets in custody.
Section \ref{sec:debt} describes exploits that would have taken advantage of the part of the bridge issuing asset representations on a destination blockchain.
Exploits that abuse the protocol's cross-chain communication system are described in Section \ref{sec:communicator}.
Finally, Section \ref{sec:interface} describes exploits that arise due to poorly defined digital assets, namely, some ERC-20 tokens.
We review related work in Section \ref{sec:related}.
Finally, Section \ref{sec:conclusion} concludes.% with some open problems and a wishlist for more secure bridges.

%% file: sections/bridges.tex
\section{Bridge Architecture}\label{sec:bridges}

We now describe the high level architecture of bridges. 
Throughout the section, we will assume that the source blockchain for an asset is Ethereum, but any blockchain that supports smart contracts can be considered without loss of generality.
We also assume that assets from Ethereum are bridged to another blockchain that supports smart contracts, which we will call ``Another Chain'' at times.

A bridge transfers assets from a \emph{source} blockchain, where the asset is originally implemented. 
The digital asset is implemented either natively, as in the case of Ether (ETH) on Ethereum, or as a smart contract.
Example of smart contract implemented tokens are ERC-20 tokens \cite{erc20} and ERC-721 tokens \cite{erc721} (i.e., NFTs).
The bridge enables unlocking or creating a representation of this asset on a \emph{destination} blockchain.

For the destination asset to be a useful representation, the bridge's representation on the destination blockchain should mimic the behaviour of the asset on the source blockchain.
In particular, the destination representation should be transferable to any party on the destination blockchain, if this is a feature of the asset on the source blockchain.
Moreover, the bridge smart contracts on the destination blockchain should accept this representation from any party on that blockchain, in order to move that asset back to the original blockchain.
This is necessary as otherwise users on the destination blockchain may not associate the representation with the original asset.
For example, if a user is given a representation of a newly minted representation of Ether on a non-Ethereum blockchain, but that representation cannot be traded for Ether on Ethereum, users are not likely to associate the same value to it or use it in the same way.

\begin{figure}
    \centering
    \includegraphics[scale=0.5]{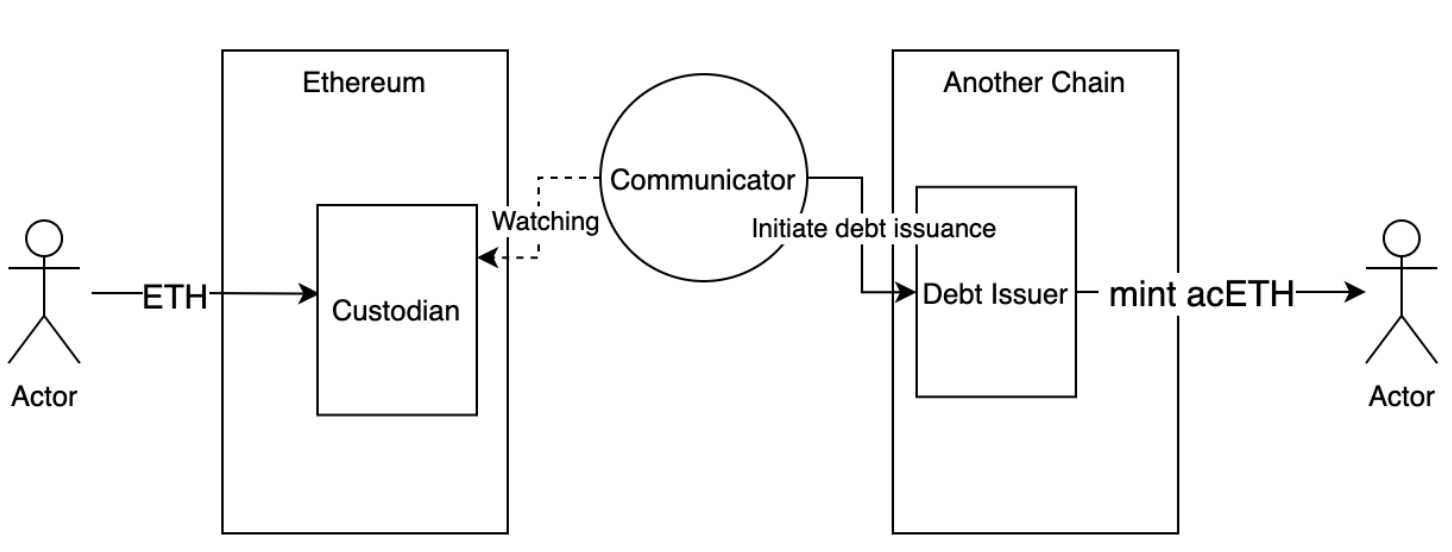}
    \caption{An actor wishes to transfer ETH from Ethereum to Another Chain. The actor sends their ETH to the bridge Custodian on Ethereum, a smart contract that accepts the asset. A Communicator waits for the Custodian to signal that it has received ETH from the user and signals the Debt Issuer on Another Chain when it detects the event. The Debt Issuer then mints acETH, the Another Chain representation of ETH.}
    \label{fig:sending-off-ethereum}
\end{figure}

We now describe how bridges work.
First, bridges have a \emph{custodian} on the source blockchain: a smart contract that locks up assets that are deposited into it.
On the destination blockchain, bridges have a \emph{debt issuer} that can \emph{create} (or \emph{mint}) digital representations of tokens for those supported by the custodian.
The custodian signals (e.g., through an Ethereum \texttt{event}) that a digital asset was received and that the corresponding debt issuer on the destination blockchain can mint a representation of the asset.
As the representation of the asset can be traded for the original asset, the representation is in fact a \emph{debt token}.
Since each blockchain is a closed ecosystem, a \emph{communicator} reads the \texttt{event} emission on Ethereum to send a signal for debt issuance on the destination blockchain.
The blockchain is a closed ecosystem because smart contracts are passive; they cannot actively (or regularly) read from non-transaction data, including data that only exists on other blockchains.
This process is illustrated in Figure \ref{fig:sending-off-ethereum}.

The custodian-debt issuer architecture is designed to avoid double spending of digital assets that have been sent across a bridge; it is important that bridges only mint digital representations only after receiving the true asset on the source blockchain. 
This prevents double spending by only having one representation of the token freely transferable at a time.

\begin{figure}
    \centering
    \includegraphics[scale=0.5]{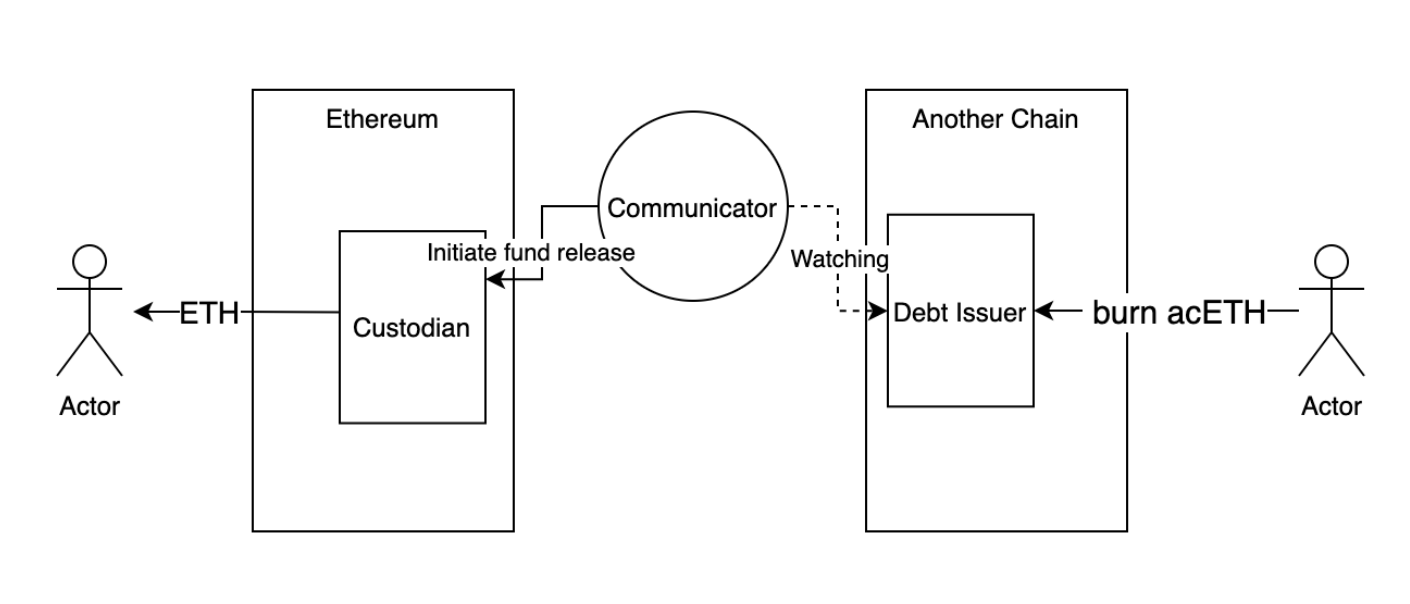}
    \caption{An actor wishes to transfer acETH from Another Chain to Ethereum. The actor sends their acETH to the Debt Issuer on Another Chain, a smart contract that accepts the asset and burns it. A Communicator waits for the Debt Issuer to signal that it has received acETH from the user and signals the Custodian on Ethereum when it detects the event. The Custodian then unlocks ETH for the user.}
    \label{fig:sending-onto-ethereum}
\end{figure}

To reverse the process, a user \emph{destroys} (or \emph{burns}) the debt token on the destination chain.
The communicator observes the destination chain, looking for every \texttt{event} corresponding to a burn. 
When the burn is complete, the communicator signals the custodian that the asset can now be released on the source blockchain.
This process is illustrated in Figure \ref{fig:sending-onto-ethereum}.

\begin{figure}
    \centering
    \includegraphics[scale=0.5]{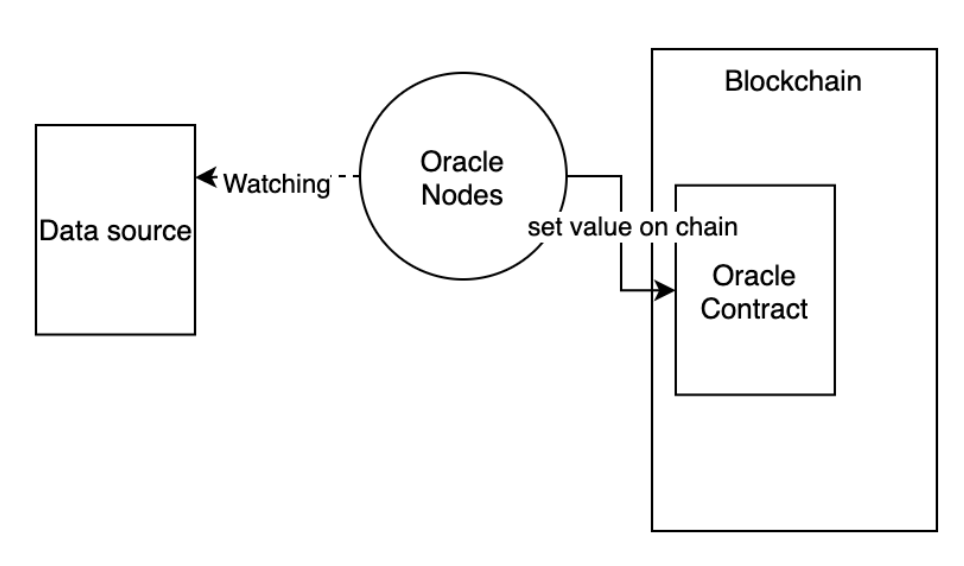}
    \caption{Oracles watch a \emph{data source} to write to an \emph{oracle smart contract} on a blockchain, to provide information that cannot be directly queried on chain. If an oracle is decentralized, it may consist of several \emph{oracle nodes}, and further, it may read from multiple data sources or write to multiple oracle contracts (not pictured).}
    \label{fig:oracles}
\end{figure}

Blockchain systems that write data to the blockchain from external sources are called \emph{oracles} (see Figure \ref{fig:oracles}).
An oracle is an agent that fetches external information into a blockchain ecosystem \cite{blockchainoracles}.
Since the communicator of a bridge writes a signal from another blockchain (from the source blockchain to the destination blockchain, or vice-versa), these components are in fact oracles.
These oracles are used to write a special signal that indicate that a transaction has been executed on another blockchain.
Thus bridges are a combination of two commonly seen structures in the blockchain space: asset custodians with debt issuers, and oracles.

There are many security considerations for the various components.
First, only messages signed by the communicator should be considered valid from the point of the custodian and debt issuer.
Otherwise, anyone can send such messages to issue debt or release assets.
Decentralised and trust-less bridges are possible \cite{DBLP:journals/corr/abs-2102-04660}, but often involve running nodes; in this way, they are similar to the blockchains that they are trying to bridge assets between.
Second, as the communicator is effectively an oracle reading from a blockchain data source, it must take care not to signal the debt issuer incorrectly. 
The communicator wait for the transaction depositing assets to be \emph{final}: guaranteed to be included in the blockchain.
A true guarantee may be impossible, as the blockchain may be re-organized.
However, if a transaction is included in a block which is included and forms the longest chain, with each block appended to that chain, the likelihood of a new chain appearing without the transaction approaches zero.
Thus the communicator should wait until the transaction depositing assets into the custodian has a sufficient number of confirmations --- blocks appended to a chain containing the deposit transaction --- following it on the source chain (or on the destination chain, if signalling to the custodian to release assets instead).

Other considerations are also important and may impact the bridge's implementation. 
Depending on the source and destination blockchains, a user may not have a one-to-one mappings of identities on both chains.
Moreover a digital asset may have different representations, each with a different implementation, on each different blockchain.

Not all blockchains have the same address format.
Bitcoin addresses are different from Ethereum addresses, but Ethereum addresses and Polygon addresses (a scaling solution for Ethereum) share address spaces.
The bridge may wish to publish a message so that anyone on a destination chain can be issued debt when assets are deposited into the custodian.
The message may be a cryptographic puzzle, so that anyone who solves it can claim it, or only be decoded by the user who deposited the asset. 
Advanced communicators may also have built-in \emph{mixers} (see e.g., \cite{beres2021blockchain}), to anonymize assets as they are transferred between chains, or other unique features.

Not all blockchains implement digital assets in the same way. 
Bitcoin uses the Unspent Transaction Output (UTXO) model, while Ethereum is account based. 
As a result, a Bitcoin that has been transferred to Ethereum will be implemented differently and have different semantics, even if it can ultimately be redeemed for a Bitcoin on the Bitcoin Network via a bridge.

Finally, since bridges rely on communicators to relay messages, it is imperative that these entities can always write on the blockchains they communicate with.
In particular, these entities should be guaranteed to be \emph{live} so that a communication between chains will always occur eventually.

Communicators may be required to collect a fee in order to guarantee this liveness.
First, bridges may be required to pay for transactions to submit transactions on either a source or destination blockchain (or both).
For example, Ethereum transactions require \emph{gas} fees, paid in Ether, which are awarded to the block producers for the chain.
This provides an incentive for the inclusion of a transaction in a block and offsets the costs of block production.
On other blockchains, another asset may be used as a gas fee, like an ERC-20 token. 
In testing or centralized solutions, gas fees may be offset by other sources (e.g., the operator of the blockchain).
Second, this fee may offset the cost of operating the communicator software.
Running a node that is required to interact with a blockchain may be non-trivial.
For example, to interface with Ethereum, an RPC endpoint is required which may be paid service or a full Ethereum node.
The former may cost per read or write to the blockchain, while the latter may be expensive to keep online and up-to-date.

%% file: sections/custodian.tex
\section{Custodian Attacks}\label{sec:custodian}

In this section, we review three exploits that have exploited the custodian component of bridges. 
The first exploit involves changing the privileged address that can access the digital assets, using cross-chain function calls.
The second exploit aims to forge proofs that are accepted by custodians to release assets.
The third exploit aims to trick the custodian into emitting deposits when it should not.

\subsection{Truncated Function Signature Hash Collisions and Forced Transaction Inclusions}\label{sec:cust-attack-1}

Depending on the structure of the bridge's custodian, privileged addresses may have access to the assets in custody.
This is common for centralized bridges, and is a requirement when only transactions from a particular whitelisted account are allowed to unlock these assets.
This requirement is the simplest way to ensure that only an appropriate communicator can unlock funds.

Moreover, if a bridge is built in a modular way, a vault holding the assets may be separate from the contracts that are written to by the communicator directly.
It may also be desirable that such a vault has its own privileged administrator.

Some bridges have the ability to execute \emph{cross-chain} function calls.
That is, the bridge can accept transactions on a source blockchain that include encodings of function calls to be executed on the destination blockchain (or vice versa).
For example, a user may wish to bridge an asset onto another chain and \emph{immediately} deposit it into a DeFi application on the destination chain.
This involves calling a deposit function on the destination blockchain, if the asset is originally on the source blockchain, or vice-versa.

The goal of this exploit is to change a bridge vault's privileged addresses to an attacker's address, using cross-chain function calls.

\begin{figure}
    \centering
    \includegraphics[scale=0.75]{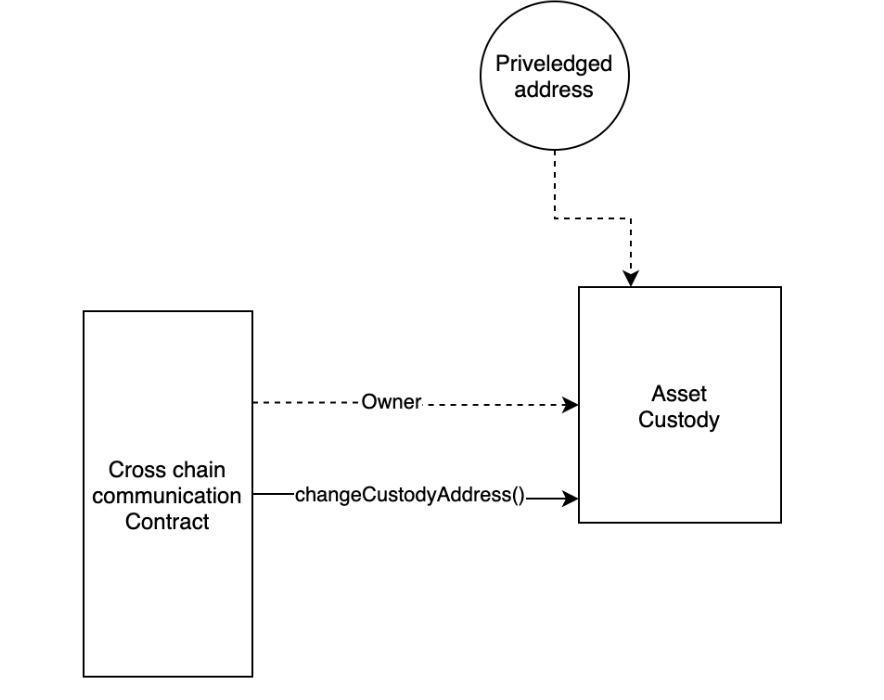}
    \caption{The structure of a custodian with additional privileged addressed to manage the custody of assets.}
    \label{fig:custodian-priv-1}
\end{figure}

The situation for this exploit is illustrated in Figure \ref{fig:custodian-priv-1}.
In this situation, the custodian has an additional field which has special roles, in addition to receiving instructions from the communicator (possibly via another smart contract).

\begin{figure}
    \centering
    \includegraphics[scale=0.5]{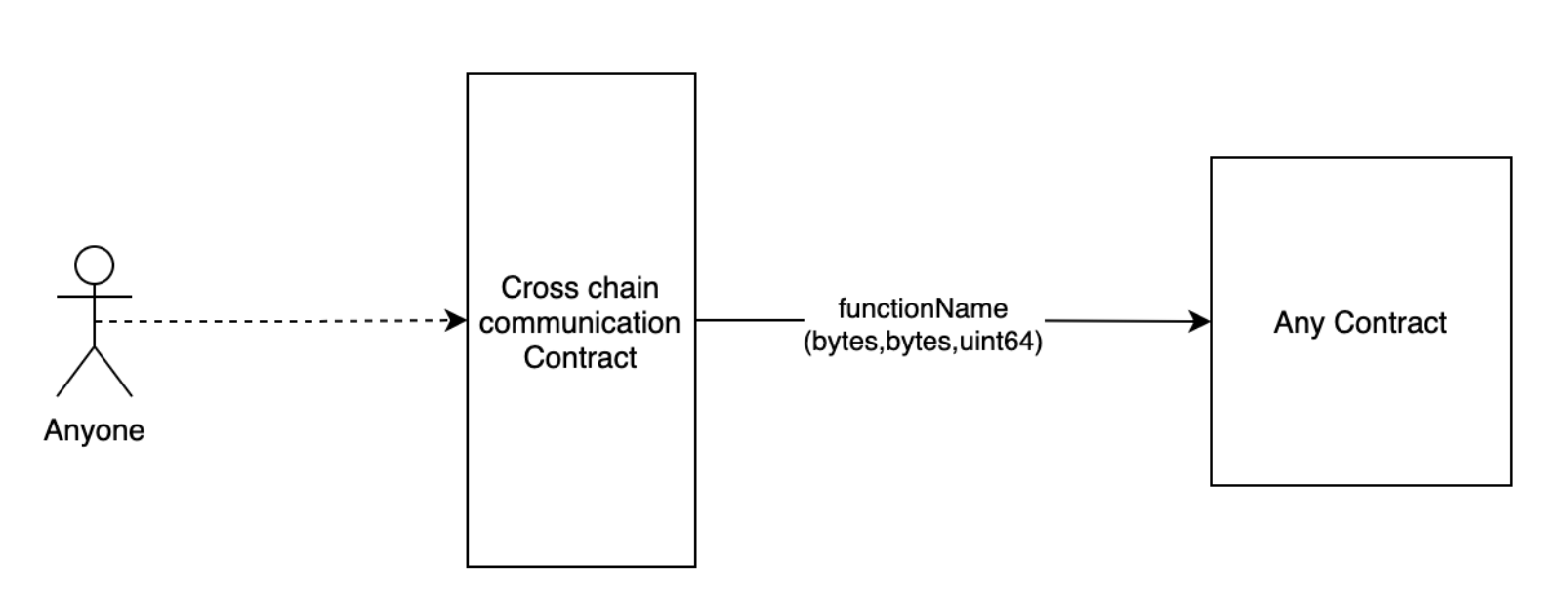}
    
    \texttt{functionName(bytes,bytes,uint256)} $\rightarrow$ \texttt{7dab77d8}
    \caption{An example of a cross-chain function call (top) and an example of the truncated hash of a function (bottom). 
    Anyone can call the cross-chain communication contract, specifying a function to call on any contract on another chain. 
    The function must be specified by a hash of the function signature: its name and ordered argument types. 
    An example is shown under the figure, where \texttt{7dab77d8} is the first four bytes of the Keccak256 \cite{keccak} hash of \texttt{functionName(bytes,bytes,uint256)}.
    This hash is used to execute a call from the other end of the bridge.}
    \label{fig:priv-attack-illustration}
\end{figure}

\begin{figure}
    \centering
    \includegraphics[scale=.65]{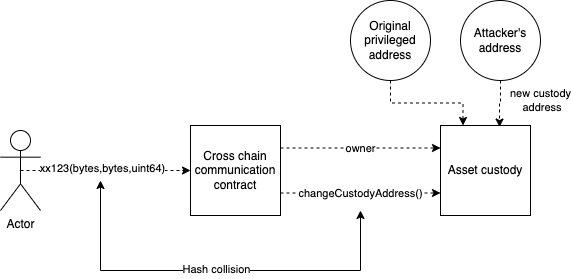}
    \caption{Illustrating which function signatures should match for a signature collision. The contract specified by the attacker should have the same signature as the function to call the method to change the custodian's privileged addresses.}
    \label{fig:priv-attack-illustration-b}
\end{figure}

With this kind of bridge structure, such an exploit occurred through the following steps, illustrated in Figures \ref{fig:priv-attack-illustration} and \ref{fig:priv-attack-illustration-b}:
\begin{enumerate}
    \item A bridge is deployed so that anyone can call its cross-chain communication contract, specifying a function to execute.
    The cross-chain function call is specified via a truncated hash of the function signature; this is common for Ethereum \cite{truncatedhash, truncatedhashB}. 
    Specifically, a function signature is the first four bytes of the Keccak256 \cite{keccak} hash of the function's name and its ordered argument types
    (Figure \ref{fig:priv-attack-illustration}).
    \item The attacker then defines a function such that, (a) the function has the same argument types, and (b) when the name along with the arguments taken by the function, the truncated hash is the same as calling a function \texttt{changeCustodyAddress} expected by the custodian. 
    The attacker specifies a contract that they own with the function signature defined above.
    Finally, the attacker specifies this function and its new contract for the cross-chain execution call, resulting in this transaction's inclusion on the destination blockchain
    (Figure \ref{fig:priv-attack-illustration-b}).
    \item The attacker notes that this transaction is now included in the destination blockchain (even if it fails), and as such can now be communicated back to the custodian on the source blockchain with proof that executing this transaction happened.
    The transaction is therefore able to be replayed on the source blockchain, where it succeeds and the attacker becomes a privileged actor of the vault.
\end{enumerate}

The sources of the error here are steps (2) and (3).
Indeed, step (2) should be nearly impossible, as hash functions are typically assumed to be \emph{collision resistant}.
A hash function is collision resistant if finding two inputs to the function that result in the same output is computationally difficult to find \cite{handbook}.
However, as the next subsection will illustrate, implementation choices made the attack feasible in one situation.
Step (3) is also a problem as a transaction's inclusion on the destination blockchain should be insufficient to replay it on the source blockchain.

%% more or less a duplciate of figure 5, but with a box around text.
%\begin{figure}
%    \centering
%    \includegraphics[scale=0.25]{figures/6.png}
%    \caption{Caption 6}
%    \label{fig:fig_6}
%\end{figure}

\subsubsection{Real World Example}

This issue was identified in the PolyNetwork bridge \cite{custodianattack1a, custodianattack1}.
Critically, the hash collision only required the first $4$ bytes of the hashes to match.
This is because the function selector, which decodes the hash, only inspects the first $4$ bytes when choosing which function to call.
Thus only a partial hash collision was necessary.
However, finding the hash collision was necessary, but not sufficient to accomplish this attack.

In practice, the attack required running a modified communicator component, the PolyNetwork \emph{Relayer}.
First, the attacker sent a transaction to the destination blockchain attempting to call the function \emph{on the destination blockchain} that is a hash collision corresponding to a legitimate transaction to change the custodian's vault owner.
This was communicated to the destination blockchain, included in the state tree, but ultimately did not execute correctly.
This was because the custodian was not on the destination blockchain, after all; it was on the source blockchain.

However, the transaction was included on the destination chain, with proof.
That is, a transaction signed by the PolyNetwork chain operator to update the vault owner was available and in the state database for the destination blockchain.
The attacker was then able to force this transaction to be executed on the \emph{source} blockchain, so that it was interacting with the custodian.
The proof was verified (since the transaction was included on the destination chain) and the transaction executed successfully (since it was now being called on the chain on which it was intended to be called).
The result was that the attacker obtained privileged roles with the vault.

\subsubsection{Solution}

The mitigation for this attack is to counter step (3).
This is because step (2) is the well-established method for resolving function signatures on Ethereum, and likely cannot be changed without introducing compatibility issues to a bridge or a hard fork of Ethereum.

A custodian should validate that the transaction, even if it originates on the destination blockchain, provided to a custodian is legitimate.
One way to do this is to ensure that the communicator cannot be bypassed.
This would prevent the inclusion proof from being considered legitimate, as the attacker would not have been able to submit it to the custodian with the communicator's signature or from the communicator's address.

\subsection{Incorrect Proof-of-Burn Verification}

Depending on the structure of a bridge's custodian, proofs are to be presented to the custodian in order to release assets.
This type of mechanism may be common for decentralized bridges, allowing anyone with a valid proof to interact with the custodian directly for withdrawals of assets, removing the need for a centralized communicator.

The goal of this exploit is to craft fraudulent proofs that would be valid for the verification process, thereby enabling seemingly correct withdrawals.

\begin{figure}
    \centering
    \includegraphics[scale=0.75]{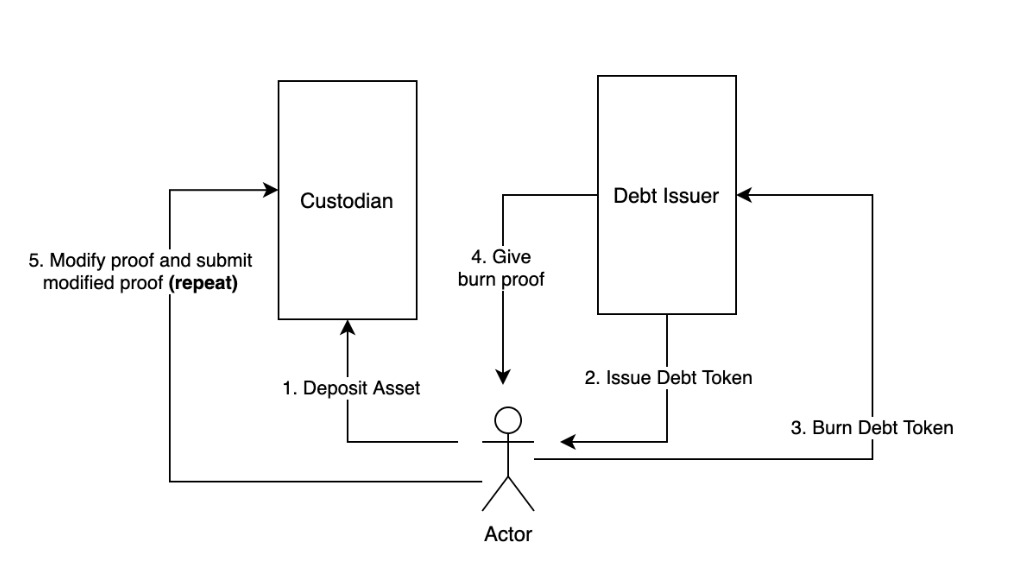}
    \caption{An exploit flow when proof-of-burn messages are not verified correctly.}
    \label{fig:proof-of-burn-exploit}
\end{figure}

This kind of exploit could have occurred through the following steps, illustrated in Figure \ref{fig:proof-of-burn-exploit}:
\begin{enumerate}
    \item An actor deposits funds into a custodian smart contract on the source blockchain.
    \item The communicator relays this information to the debt issuer on the destination blockchain for the bridge, and the debt issuer provides the actor with a debt token.
    \item The actor burns the debt token by depositing it back into the debt issuer.
    \item The actor receives a so-called \emph{proof-of-burn} for the token. The proof-of-burn is a string generated by the debt issuer showing that the debt token was burned.
    \item The actor submits a (modified) proof-of-burn to the custodian, to unlock assets on the source blockchain, and the custodian considers the proof valid.
\end{enumerate}

The source of the error here is step (5), which enables an attacker to submit a modified proof-of-burn (alongside the original proof-of-burn) to withdraw funds.
The real-world exploit occurred because the proof had a leading byte that was not verified by the custodian when releasing funds, which we now describe.

\subsubsection{Real World Example}
This exploit was detected and patched on the Polygon/Matic Ethereum-Plasma bridge before any harm could be done \cite{polygondoublespendhack}.
We outline the specifics of how this particular issue manifested itself for completeness, though this type of exploit may have different manifestations depending on the (incorrect) implementation of proof generation and verification for a particular bridge.

In this case, the custodian is to release funds if a proof-of-burn for the debt token is specified in a particular Merkle Patricia trie (see e.g., \cite{DBLP:journals/corr/abs-2106-04826}) representing the state of the destination blockchain.
In this case, the proof-of-burn includes a path to the leaf in the Merkle Patricia trie which specified that that the debt token was burned (the transaction should be included when the actor submits it on the destination chain).
This proof-of-burn included a \texttt{branchMask} parameter that should be unique. 

The \texttt{branchMask} parameter is encoded with so-called \emph{hex-prefix} encoding~\cite{ethereum}. 
But at some points within the system implementation, the parameter is encoded and decoded into 256-bit unsigned integers, and during this process some information is lost.
In particular, a path in the Merkle Patricia trie may have multiple valid encodings with the system.

The system was implemented to determine the path (in a trie) length encoded by a hex-prefix encoding.
To use the encoding's length, it is important to know if the length of the path is even or odd; this affects how the encoding is later expanded.
The system was implemented to check that the parameter's first nibble (4 bits) represented $1$ or $3$; if so, it considered the path length to be an odd number.
However, it was also implemented such that in the event that the first nibble is not $1$ or $3$, the first byte (8 bits) is discarded but verification proceeds.
Thus, there are $2^8-2(2^4)=224$ possible ways to encode a path in the Merkle Patricia trie in the situation where the first byte is discarded.
In particular, there are $2^8$ encodings for every possible bit setting of the first byte, minus the cases where the first nibble is either $1$ ($2^4$ cases; every configuration of the last 4 bits) or $3$ (also $2^4$ cases for each configuration of the last 4 bits).

Thus the attacker would simply find a valid proof where the initial nibble was not $1$ or $3$, use it, and then replay the transaction for each of the remaining $223$ combinations of bits for the first byte.
In each case, the proof-of-burn would look legitimate and the exit would succeed, subject to delays in confirmations, delay periods, or other specific requirements of this bridge and the blockchains it connected.

\subsubsection{Solution}
The remedy for this exploit is correct implementation of proof verification.
The original reporter of the issue notes that the first byte should in fact always be zero, reducing the number of times a valid proof-of-burn can be used to only once \cite{polygondoublespendhack}.

If the relevant proofs are built and verified correctly, this exploit will not be common, subject to common cryptography assumptions like the collision resistance of hash functions and the inability to forge digital signatures.

\subsection{Inconsistent Deposit Logic}

Bridges are often built for custom blockchains.
For example, anyone developing a rollup may have a token that is used for governance or to be used as payment for gas on the rollup (instead of ETH).
As a result, sometimes bridges have custom functionality for some tokens.

Moreover, a token can be ``wrapped'' within another token.
Most commonly, Ether (ETH) is often wrapped into \emph{wrapped Ether} (wETH).
This is helpful because some decentralized applications do not wish to treat Ether differently from ERC-20 tokens, and wrapped Ether is an ERC-20 token.
This can be helpful, as native ETH lacks a \texttt{transferFrom} function, among other helpful functions that are available to ERC-20 tokens.
As a result, there is a wrapped Ether smart contract on the source blockchain that essentially lets anyone lock one ETH to mint one wETH.

The goal of this exploit is to trick the custodian into emitting events for deposits which are not real.

%% Duplicate of 10, which is slightly more readible.
%\begin{figure}
%    \centering
%    \includegraphics[scale=0.25]{figures/9.png}
%    \caption{Caption 9}
%    \label{fig:fig_9}
%\end{figure}

\begin{figure}
    \centering
    \includegraphics[scale=0.65]{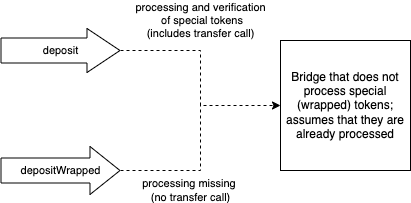}
    \caption{Two separate paths to deposit into a custodian contract.}
    \label{fig:cust-attack-3}
\end{figure}

This kind of exploit occurred through the following steps, illustrated in Figure \ref{fig:cust-attack-3}.
It is fairly restricted in scope and requires special tokens, like wrapped Ether, to be handled differently than unwrapped assets.

\begin{enumerate}
    \item The bridge is established in such a way that its final logic for emitting deposit events is after processing of wrapped assets. 
    The bridge is also (incorrectly) built so that unwrapped assets allow this logic to be called, without actually supporting the transfer of those assets.
    \item An attacker deposits assets into the custodian, without first wrapping the assets.
\end{enumerate}

The second step is the source of the issue, and is exemplified in the real-world manifestation we now describe.

\subsubsection{Real World Example}

This error occurred for the Meter bridge \cite{custodianattack3}.
In this occurrence, the bridge expected all assets to be transferred in a wrapped form, and assumed a deposit of unwrapped assets is a mistake.
However, the deposit of unwrapped assets was encoded within the same event logic that accepts wrapped assets, even though unwrapped assets were not accepted by the custodian.
As a~result, the custodian still emitted an event saying that funds had been transferred, even though the custodian never received them.
That is, the caller continued to own their assets, but the custodian still emitted an event.

\subsubsection{Solution}

This particular attack is not conceptually involved. 
Its mere existence was enabled by a bug in the code and the branching logic. 
It serves as a reminder to the bridge developers that Ethereum's native Ether is not an ERC-20 token, and both the cases of transferring Ether and its wrapped form need to be handled properly. 
Good engineering practices, including implementing tests, should suffice to mitigate the problem in the future.

%% file: sections/debt-issuer.tex
\section{Debt Issuer Attacks}\label{sec:debt}

In this section we review one exploit on the debt issuer component of a bridge.
The exploit aims to arbitrarily mint debt tokens on the destination blockchain.

\subsection{Bypassing Signature Verification}

The exploit aims to arbitrarily mint debt tokens on the destination blockchain.
In doing so, the attacker can trade these tokens back in, honestly, and receive the corresponding assets on the source blockchain, as long as such assets are available, or for other assets on the destination blockchain.

\begin{figure}
    \centering
    \includegraphics[scale=0.75]{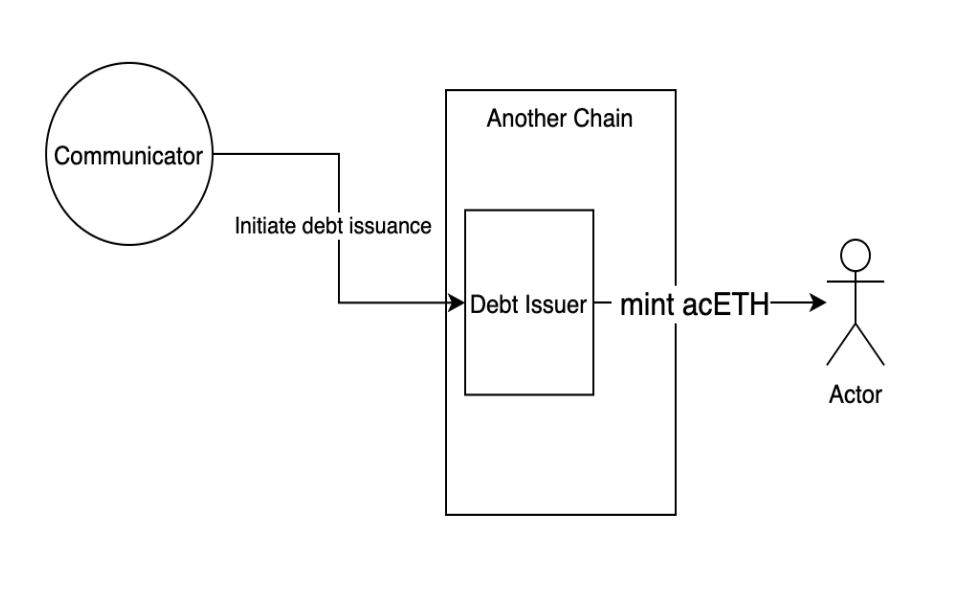}
    \caption{Components required for debt token issuance.}
    \label{fig:debt-issuer-components}
\end{figure}

Recall that the debt issuer smart contracts live on destination blockchain, mint debt tokens which are representations of assets on the source blockchain, and receive minting signals from a communicator (see Figure \ref{fig:debt-issuer-components}).

\begin{figure}
    \centering
    \includegraphics[scale=0.5]{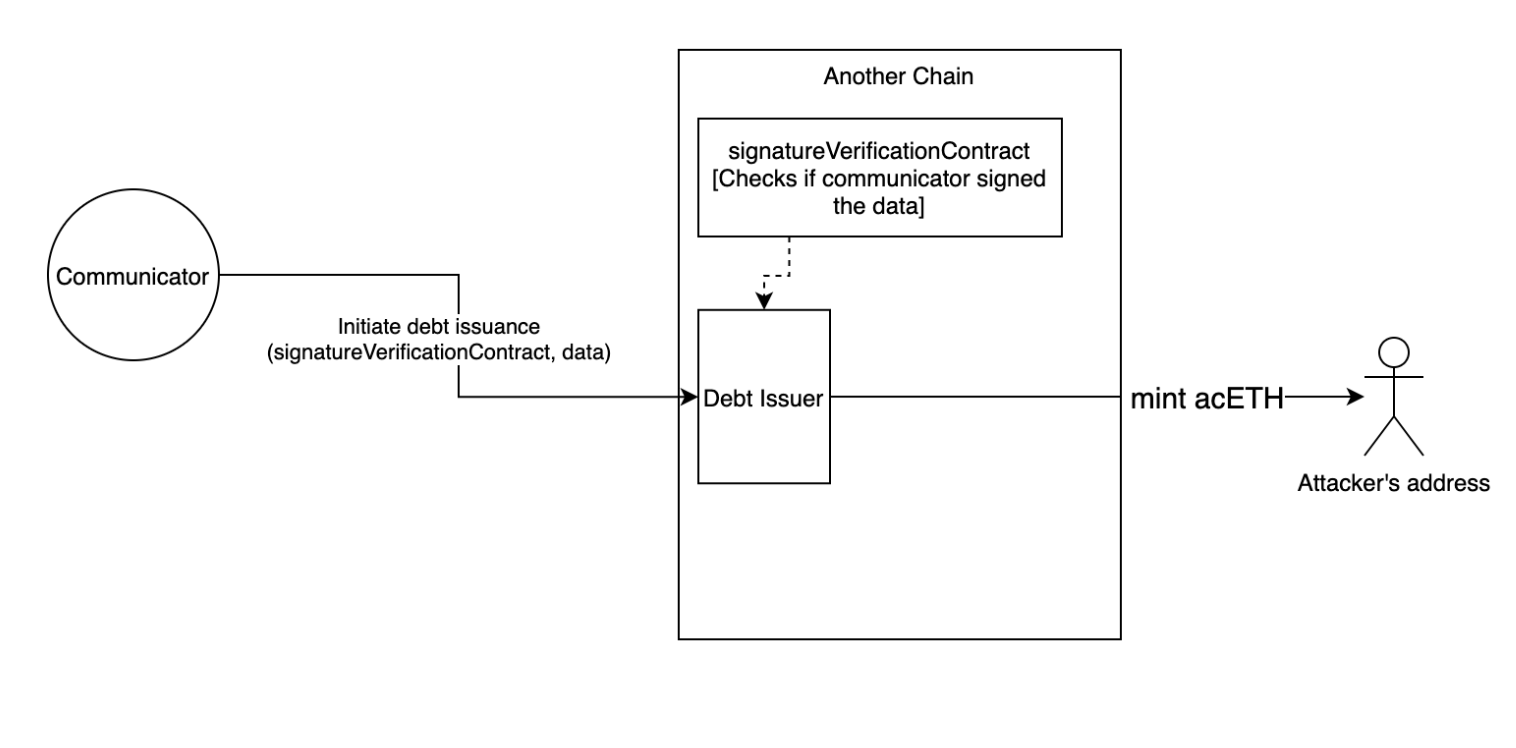}
    \caption{Modularized signature verification logic in a debt issuer.}
    \label{fig:attack-on-debt-issuer-1}
\end{figure}

In the most straightforward implementation of debt issuers, these components mint tokens only after receiving a signed message from a communicator.
This prevents unwanted tokens from being minted on the destination blockchain.
To check the validity of such a signature, verification logic may be placed in a smart contract which is external to the contract issuing the debt tokens on the destination chain (see Figure \ref{fig:attack-on-debt-issuer-1} for an example).
Moreover, if there are several communicators in a bridge, each might have its own verification logic, and modularising this logic may make sense from an engineering standpoint.
This would enable verification of signatures from multiple sources, each with its own verification scheme.
When a message is received in this situation, it could therefore include the address of the verification logic to be used.
The logic for determining which verification contract should be used must be matched to the message, and including the address of a contract that implements the verification logic is a straightforward implementation.
However, problems arise if matching allows messages to be matched with arbitrary verification logic, as we now describe.

\begin{figure}
    \centering
    \includegraphics[scale=0.5]{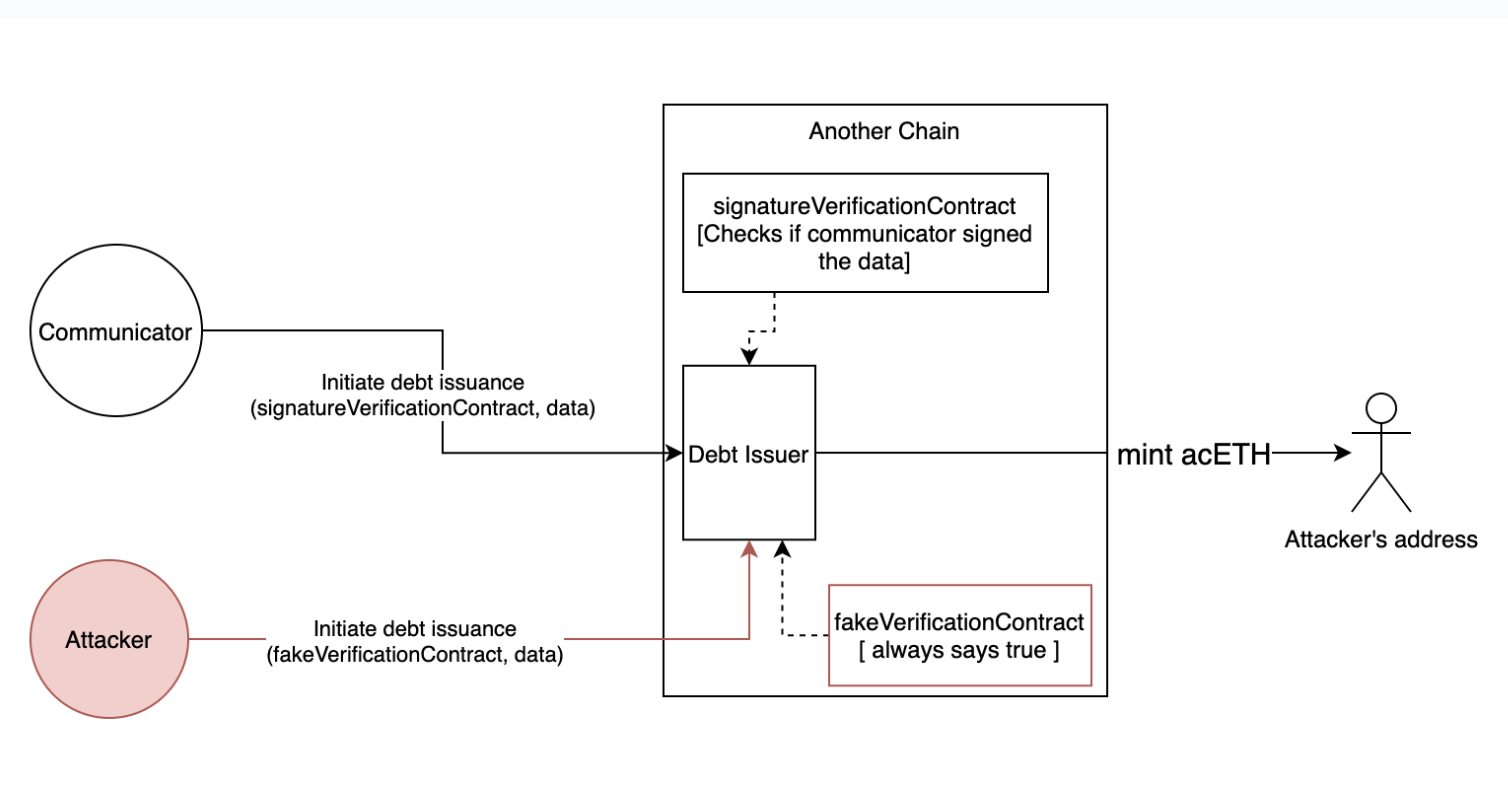}
    \caption{Changing the signature verification logic in a debt issuer to mint debt tokens.}
    \label{fig:attack-on-debt-issuer-2}
\end{figure}

This exploit was executed using the following steps, illustrated in Figure~\ref{fig:attack-on-debt-issuer-2}:

\begin{enumerate}
    \item An attacker deploys a smart contract on the destination blockchain that has a function that the debt issuer expects to call to verify a signature.
    The function is implemented so that any signature is ``verified", possibly by implementing the verification function so that it always returns \texttt{true}.
    This verification contract therefore accepts any string as a valid signature.
    \item The attacker from step (1) sends a debt issuance signal to the debt issuer, referencing the smart contract they deployed in step (1) as the verification logic.
    \item The debt issuer provides debt tokens to the attacker.
\end{enumerate}

The source of the issue here is in steps (2) and (3).
The debt issuer should not have accepted just any smart contract as verification logic.
After the attacker gains the debt tokens, they can behave honestly to bridge the assets back to the source blockchain, stealing funds from the custodian.

\subsubsection{Real World Example}\label{sec:first-deb-attack}

Unfortunately, this situation was exploited on the Wormhole bridge \cite{wormhole}.
The result was that about 120,000 Ether was minted on Solana, which was worth about \$323 million USD at the time the exploit occurred.
Much of this Ether was transferred back to Ethereum.

\subsubsection{Solution}

The example in Section \ref{sec:first-deb-attack} was enabled by the attacker's ability to provide both the signature (used to confirm the authenticity of the transaction) and the reference to the signature verifier (used to confirm the signer's authorization to issue the transaction) within the user transaction. 
As a result, the attacker was able to authorize any calls via a friendly custom verifier. 
Therefore, a clear prevention of the attack is ensuring that verifiers cannot be provided by users. 
Verifiers need to be absolutely trusted elements of the system, and as such, can be deployed only by trusted entities, and users cannot be provided with an option to choose a dishonest verifier to authorize their transaction. 

%% file: sections/communicator.tex
\section{Communicator Attacks}\label{sec:communicator}

In this section we review two exploits targeting the communicator component of a bridge.
The first exploit aims to trick the communicator into forwarding invalid messages from one blockchain to the next, while the second uses a 51\% attack on a blockchain to cause a blockchain re-organization after the communicator receives a valid message.
These exploits can be thought of as polluting the data source of an oracle,  the communicator.

\subsection{Forwarding Invalid Messages}

The goal of this exploit is to trick the communicator into forwarding invalid messages from the source blockchain. 
This will result in incorrect debt issuance on the destination blockchain, minting debt tokens that are not mapped to assets in custody on the source blockchain.

\begin{figure}
    \centering
    \includegraphics[scale=0.75]{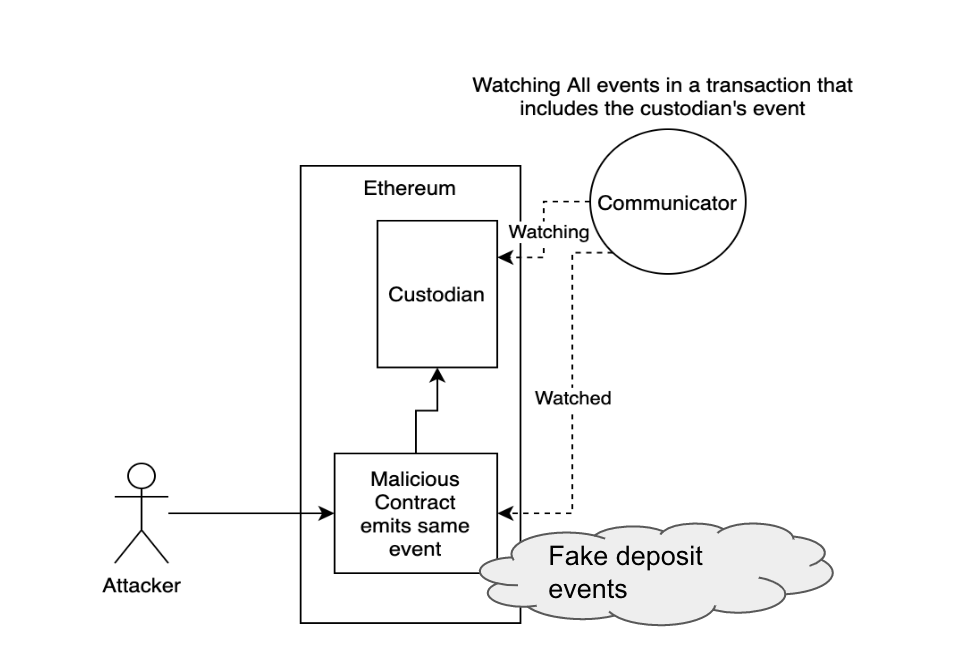}
    \caption{An illustration of using fake events on the Ethereum blockchain to exploit a naive bridge communicator.}
    \label{fig:communicator-attack-1}
\end{figure}

The exploit proceeded according to the following steps, illustrated in Figure~\ref{fig:communicator-attack-1}:
\begin{enumerate}
    \item The bridge is established in such a way that its communicator watches events emitted from the source blockchain. 
    The communicator watches for these events on transactions that deal with a particular address, namely, the address of the custodian for the bridge.
    Notably, it watches all events such a transaction.
    \item An attacker creates a smart contract on the source blockchain.
    This smart contract has a method to interact (correctly) with the custodian address, but also emits an event that is identical to the one emitted by the custodian smart contract, immediately before or after interacting with the custodian smart contract.
    The event emitted by the attacker's contract contains parameters that appear correct to the communicator.
    \item The attacker interacts with the custodian (correctly), via the smart contract deployed in step (2).
\end{enumerate}

The source of the error here is step (1), which has an incorrect implementation of a communicator.
In particular, the communicator watches for \emph{all} events on a transaction that interacts with the custodian's smart contract, and parses them if they look legitimate.
In turn, it sends signals to the debt issuer for each event it detected.
However, because the communicator watched all events in the same transaction, regardless of which contract emitted them, the events emitted by the contract deployed by the attacker in step (2) are also parsed by the communicator.
The result is that the debt issuer mints more debt tokens than exist in the custodian, and the bridge has failed to faithfully map the digital asset.

\subsubsection{Real World Example}

This exploit occurred with the pNetwork pBTC-on-BSC, that is, the pNetwork-wrapped Bitcoin (pBTC) on Binance Smart Chain bridge \cite{pnetwork}.
The bridge operators themselves, pNetwork, examined the impact and determined that both legitimate and fraudulent logs for withdrawal requests by the communicator were processed due to a bug in the code.
The bridge execution was paused as the attack was detected to minimize loss of funds.

\subsubsection{Solution}
A communicator watching events is a core part of the off-chain application logic of the bridge. 
This can have many forms, depending on the specific implementation of the communicator.
In order to mitigate this attack, the developers of the communicator have to ensure that only the events emitted by the custodian smart contract are watched and acted upon. 
This may be a non-trivial task as event logs in the Ethereum protocol are not equipped with any cryptographic means of authentication (such as digital signatures).
Moreover, these event logs are organized and associated to blocks quite specifically using so-called \emph{Bloom filters} \cite{Bloom70} so that they remain quickly accessible when searching the blocks (see~\cite{ethereum} for details). 
A thorough understanding of the stack and libraries used for developing the communicator is crucial in order to ensure that only the relevant events that cannot be spoofed are taken into account before issuing assets on the destination chain. 
This is a generally applicable rule for all privileged activities that the communicator performs based on events that it listens to, i.e., it may apply to much more than just issuing assets.

\subsection{Short Term 51\% Attacks on the Source Blockchain}

The goal of this exploit is to trick the communicator into forwarding messages that will disappear after a re-organization of the source blockchain. 
Again, this will result in incorrect debt issuance on the destination blockchain, minting debt tokens that are not mapped to assets in custody on the source blockchain.

A blockchain is rarely a real ``chain.'' 
Instead, actors who propose new blocks onto a blockchain are competing for their block to be included.
In a so-called \emph{proof-of-work} (see e.g., \cite{DBLP:journals/comsur/XiaoZLH20}) system like Ethereum or Bitcoin, proposers must solve a cryptographic puzzle for this right; proposers who solve this problem are \emph{miners}.
Moreover, rules like heaviest computation dictate which blocks should be considered the so-called \emph{canonical chain}, onto which new blocks should be appended \cite{qsbook}.
Along the way, forking of the chain occurs and some blocks are orphaned from the canonical chain.
However, colluding actors can influence which chain is satisfies the rules to consider a chain canonical, if they have sufficient computational power.
This is the basis of a so-called 51\% attack, as it requires a majority of the computational power (among all miners) to execute.

A 51\% attack is typically expensive. 
However, the cost of the attack is tied to the computational power in the network and the duration of the attack.
More computational power and a longer attack duration increase the cost, while shorter attacks are cheaper.
For blockchains which use different methods to append blocks, like those which use so-called \emph{proof-of-stake} (see e.g., \cite{DBLP:journals/comsur/XiaoZLH20}) consensus algorithms, the cost of this attack will depend different factors.
For example, proof-of-stake systems may have an increased risk for this this attack if there are too few validators who propose blocks, the validators can be bribed or collude, or if the cost required by stakers is too low.

The exploit would proceed according to the following steps:
\begin{enumerate}
    \item An attacker honestly deposits funds into a bridge via the source blockchain's custodian, which issues debt on the destination blockchain.
    \item After waiting a small-but-not-too small amount of time (enough for several confirmations of the transaction, perhaps about 5; this is about 15 additional blocks on Ethereum at the time of writing), the attacker rents computational power to enact a moderately long 51\% attack (about 1 hour). 
    This attack establishes another chain which is canonical after the attack but does not have the attackers transaction from step (1) included on the chain.
\end{enumerate}

The source of the exploit here is contained in step (2), where the attacker re-organizes the source blockchain but also has the issued funds on the destination blockchain.
This attack is a specific instance of double-spending a token \cite{DBLP:journals/access/IqbalM21}.

\subsubsection{Real World Example}
This exploit has not yet been executed (on Ethereum). 
At the time of writing, the website \texttt{Crypto51.app} \cite{crypto51} reports that the cost of a 1-hour long 51\% attack on Ethereum would cost \$600,374 USD. This value is lower for chains that have less computational power associated with them. Nevertheless, as the real-world examples of bridge attacks referenced in this paper demonstrate, the attacker's profit often exceeds this number.

\subsubsection{Solution}

The inherent cause of the attack is the bridge's wrong assumptions about the finality of blocks. The bridge needs to ensure that if a reorganization of the source chain happens and the deposit transaction becomes invalidated, the same invalidation happens on the target chain. 
This is a difficult task for bridges that are not implemented and operated natively by the target blockchain itself, and reside on it in the form of a third party application. The native bridges may implement a mechanism that keeps track of deposit nonces and the total bridged value on the source chain, and subsequently require ``commiting'' the nonce and value sequence in the transactions that release the assets from the custodian. 
The nonces would have to have a fixed sequence that prohibits skipping (e.g., integers that increment by 1 with every deposit) so that they guarantee that if a deposit transaction is dropped, or it value changes, all the subsequent deposits to the target chain and withdrawals from it become invalid (i.e., result in failed transactions).
Consequently, the bridge would need to ensure that such a reorganization is properly reflected on the target chain so that the users whose assets were now not deposited to the target chain or withdrawn have their balances adjusted accordingly, and the integrity of the asset amounts between the two chains is not violated.

It is is also important to note that the roles of a source and target chain are to a certain extent symmetric---a chain that is in the role of a source during the deposit may be in the role of a target during the withdrawal. While a bridge may be a native component of one of the chains and may ensure that the chain can respond to the reorganization of another chain, it is unlikely that it would be able to guarantee such a reorganization for both the chains, in particular, for the Ethereum mainnet. 

The prevention of the attack described in this section is a difficult problem and it would make for a great subject for future research. 

%% file: sections/interface.tex
\section{Token Interface Attacks}\label{sec:interface}

In this section we describe some exploits based on the token interfaces used in bridges.
The first exploit relates to token approvals for bridges, while the second exploits the EIP-2612 \cite{erc20permit} interface function built into some ERC-20 tokens.

\subsection{Infinite Approvals and Arbitrary Executions}\label{sec:token-attack-1}

The goal of this exploit is to take user's funds directly, rather than stealing them from the custodian or debt issuer, by leveraging bridge components which can call other smart contracts.

A valuable use case of ERC-20 tokens is the ability to approve others to spend your tokens.
For example, you may wish to approve a bridge to spend your tokens, so that in the future if you use a decentralized application to interact with the bridge, it can take funds on your behalf, through the decentralized application you are interacting with.
This is achieved by having the user call \texttt{approve} (possibly specifying some specific amount) on the token, listing the bridge's relevant smart contract address, and later having the bridge call \texttt{transferFrom} to take funds from the user. 
The latter call will only succeed if the user has approved the bridge to act on the user's behalf.

Due to large gas concerns, users often grant applications and bridges \emph{infinite} approvals.
This is because the \texttt{approve} call is a transaction that must be executed on chain, for which gas must be paid.
As a result, an infinite approval removes the requirement of subsequent approval calls, saving the user gas fees in the future.

Moreover, recall from Section \ref{sec:cust-attack-1}, due to the composability of smart contracts (especially popular in DeFi applications), bridges often call other smart contracts directly.
To do this, a bridge may have an arbitrary function \texttt{execute} which takes an ABI encoded description of the function to call (see also Section \ref{sec:cust-attack-1} and \cite{truncatedhashB}).

\begin{figure}
    \centering
    \includegraphics[scale=0.85]{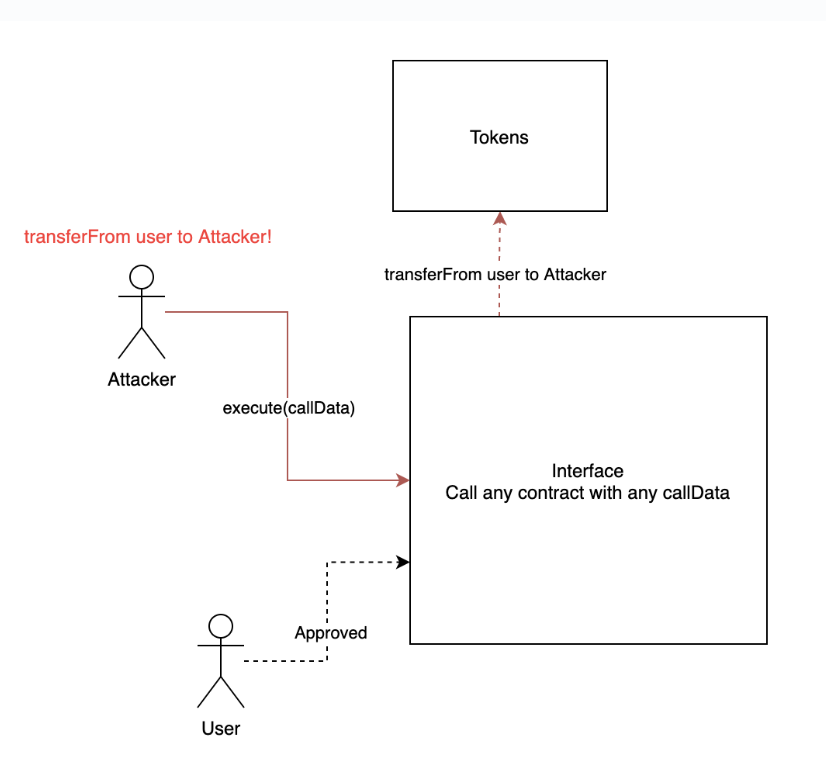}
    \caption{An illustration of exploiting bridges with arbitrary execution with infinite approvals from users.}
    \label{fig:attack-on-token-interface-1}
\end{figure}

This exploit proceeded according to the following steps, illustrated in Figure \ref{fig:attack-on-token-interface-1}:
\begin{enumerate}
    \item A user provides a bridge that can call smart contracts with an infinite approval to a token.
    \item An attacker calls \texttt{execute} with an encoding of \texttt{transferFrom} to take the honest user's tokens, rather than their own.
    This succeeds since the \emph{bridge} executes the \texttt{transferFrom} call, and it has approval to take the user's tokens.
    However, since the attacker initiated the call, the debt issuer on the destination chain issues the debt in the name of the attacker.
\end{enumerate}

The source of the error for this exploit is in step (2), in which the debt is incorrectly issued to the wrong party.
After the attacker receives the debt token on the destination blockchain, they can bridge the asset back to the source blockchain and withdraw the funds.
The user is now powerless to recover those tokens.

\subsubsection{Real World Example}

This exploit was possible on an earlier version of the Multichain (formerly ``AnySwap'') project \cite{interface1}.
The attack vector was documented before it was exploited, and the finders were awarded a \$1,000,000 USD bounty for finding and reporting the issue, which they first demonstrated on a local fork of Ethereum.
At the time the exploit was reported, almost 5,000 accounts had granted infinite approval to the bridge in question.

\subsubsection{Solution}

As the vulnerability is enabled by the bridge issuing debt tokens to a user who did not supply the tokens on the source chain, one possible remedy is to ensure that the debt is always issued to the account that provided the tokens on the source chain. 
However, this strongly limits the design of the bridge and may cause problems when bridging tokens in the custody of a smart contract. 
A specific concern with this solution is the use of a so-called \emph{multisig wallet} (see e.g., \cite{DBLP:conf/asiacrypt/BonehDN18}) --- a smart contract that holds tokens on behalf of multiple users whose joint signatures are required for releasing such tokens.
Such a smart contract may be available on the source chain, however, due to how the smart contract addresses are determined (see~\cite{ethereum}), such a multisig wallet may not be available on the target chain. 
Other measures, such as disallowing generic calls to functions such as \texttt{execute}, may negatively impact the required features of the bridge, and thus do not appear viable without disrupting the business logic.

\subsection{Permits and Non-Reverting Fallback Functions}

Similar to Section \ref{sec:token-attack-1}, the goal of this exploit is to take user's tokens directly, rather than stealing them from the custodian or debt issuer, by leveraging bridge components which can call functions in poorly implemented token contracts.

Some ERC-20 tokens have a \texttt{permit} function, which enables a user to sign a message enabling others to use one's tokens; these implement EIP-2612 \cite{erc20permit}. 
A message is signed using the account's private key.
These messages are not transactions (they are not executed on-chain), and do not use gas; as a result, they are attractive in some settings as they are free for users.
Once someone has obtained a signed \texttt{permit} message, they can go to the smart contract for the token and call a function to verify the signed message.
If the verification succeeds -- that is, a \texttt{verify} function for the \texttt{permit} \emph{does not revert} -- the holder of the permit is approved to use the signer's tokens (e.g., via the \texttt{transferFrom} function).
A function \emph{reverts} on Ethereum if it runs out of gas, an error occurs, or an assertion (written as a \texttt{require} or \texttt{assert}) fails in the code being executed.
The permit holder can obtain the approval by calling a function \texttt{redeemPermit} and ``using up'' the permit.
The fact that the verify permit function is expected to revert if the verification fails is key, as we explain next.

The expectation that a function reverts is problematic when developers are not aware of the expectation. 
Smart contracts on Ethereum have a \texttt{fallback} function, which is called whenever a function that is supposed to be called on a smart contract cannot be found (i.e., it is not implemented).
If an ERC-20 smart contract implements a fallback function that does not revert, every time a function that does not exist within the implementation is called, the call will succeed.
This can be problematic, as we now exemplify in this exploit.

\begin{figure}
    \centering
    \includegraphics[scale=0.75]{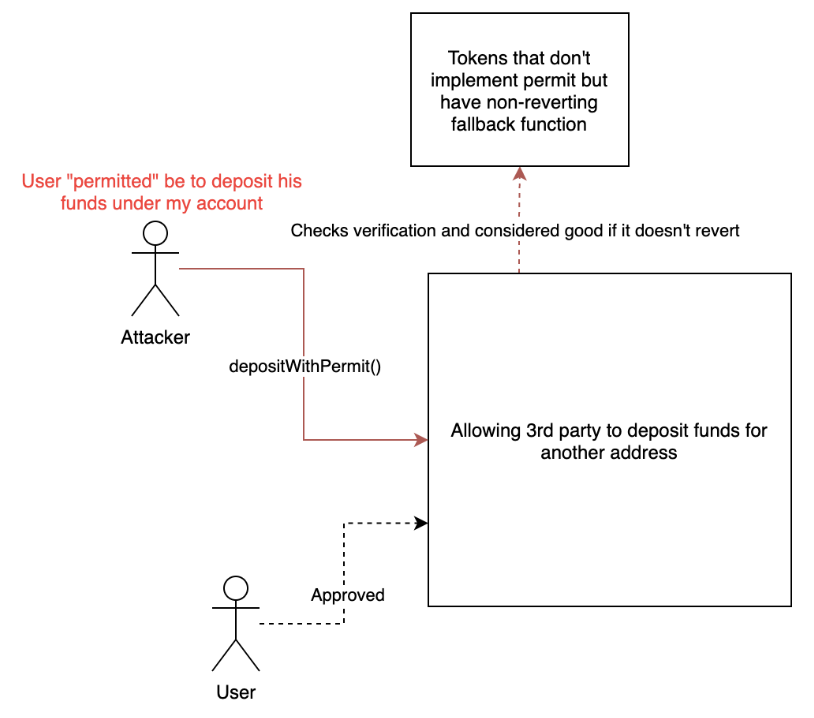}
    \caption{An illustration of exploiting bridges supporting tokens with the \texttt{permit} functionality.}
    \label{fig:attack-on-token-interface-2}
\end{figure}

The exploit proceeded according to the following steps, illustrated in Figure \ref{fig:attack-on-token-interface-2}:
\begin{enumerate}
    \item A user wishes to bridge tokens to another chain from Ethereum where (a) the bridge supports \texttt{permit} redemption for tokens, and (b) the bridge has custody of at least one token that does not implement the EIP-2612 \texttt{permit} functions but implements a \texttt{fallback} function that does not revert.
    \item The user gives the bridge infinite approvals (and may or may not successfully send some tokens over the bridge).
    \item An attacker asks the bridge to redeem a string, claiming it is a permit, for approval on the token used by the honest user in steps (1) and (2).
    The bridge attempts to verify the supplied string; as the \texttt{verify} function is not implemented for any \texttt{permit} but the \texttt{fallback} function never reverts, the bridge accepts the permit. 
    Next, the bridge contract calls the \texttt{redeem} function for the \texttt{permit}, which is not implemented either, but since the \texttt{fallback} function never reverts, the bridge thinks it has succeeded in obtaining the approval.
    The bridge calls \texttt{transferFrom} on behalf of the attacker onto the bridge, and issues debt in the attacker's name.
\end{enumerate}

As in the Section \ref{sec:token-attack-1}, the source of the error for this exploit is in step (3), in which the debt is incorrectly issued to the wrong party. 
After the attacker receives the debt token on the destination blockchain, they can bridge the asset back to the source blockchain and withdraw the funds.
However, this reasoning is different: the bridge was not implemented to check that \texttt{verify} was actually called, rather than the \texttt{fallback} function.

\subsubsection{Real World Example}
The situation above was described as a feasible bug in the deprecated Polygon Bridge Zap \cite{interface2}.
The operators fixed the issue in subsequent versions even before it was discovered, but because a version was already deployed on the blockchain the exploit happened.
To mitigate this, the bridge operators submitted transactions to withdraw funds themselves, but were front-run by an arbitrage bot.
However, the arbitrage bot later returned the front-run profits when the bot operator learned that the transaction was executed with good intention.
The vast majority of funds were held in escrow by the Polygon team after this effort, resulting in no significant loss of funds from the bridge.

\subsubsection{Solution}

The core idea of the vulnerability lies in the fact that the bridge calls a non-existent function on a token to redeem the permit, and that the token's implementation allows calls to non-existent function without reverting. 
In order to eliminate it, the bridge needs to break this condition. This means that the bridge needs to be aware of the implementation of the token, and avoid attempts to use the permit logic if the token does not implement it, or does not implement it correctly. 
As the Ethereum blockchain does not allow for checking interface and implementation on-chain, a list of tokens maintained by the operator of the bridge or an off-chain logic would have to be used for detecting whether the permit mechanism should be available.

%% file: sections/related.tex
\section{Related and Future Work}\label{sec:related}

To our knowledge, there is no systematization of the attacks that have occurred on bridges in recent years.
Others have studied these components, they largely do so from a theoretical point of view, rather than reviewing the faults of previous systems.
We first list a few of the relevant works.

McCorry et al.~\cite{cryptoeprint:2021/1589} review the literature involving bridges and provide a detailed breakdown of roles and components.
Although their terminology differs from ours, the key concepts they identify for bridges are compatible with the major components we define in Section \ref{sec:bridges}.
Their work provides a more fine-grained overview; for example, they dive into concepts like the operator of a communicator (i.e., whether it is centralized, involves a multi-signature wallet, or purely trustless), protocol assumptions (e.g., expanding properties beyond liveness), and things like rate-limiting transactions for the bridge.
They offer research directions for improvements based on various assumptions and dilemmas that bridges aim to solve or sidestep.
However, their threat models are high level and theoretical; they are not reviews of attacks that have occurred in practice.
For example, they emphasize the need to prevent censorship of communicators.
This is critical for bridge design and complementary to the review of concrete issues we review in this work.
This is discussed somewhat less formally for layer two solutions in \cite{l2beatbridges}.

Zamyatin et al.~\cite{cryptoeprint:2019/1128} studies communication between blockchains, necessary for the communicator component of a bridge.
They look at which assumptions are required, classify and evaluate existing cross-chain communication protocols, and generalise the systems we describe in this work.
They list challenges that must be overcome for safe and effective cross-chain communication.
These challenges show that bridge communication is difficult and indicate why bridges are so complex and therefore prone to implementation errors.

For future work, it would be interesting to study preventative measures for these attacks.
While many attacks presented are implementation specifics, we wonder if a general framework or set of standards can help mitigate these issues.
In particular, custodian or debt issuer standards may reduce errors with cross-chain calls or decrease erroneous event emissions.
Such a standard could be an interface akin to the ERC-20 or ERC-721 standards.

Moreover, a wishlist of specific properties for security should be explicated.
The related work targets the high level properties, but is insufficient to guide new bridge developers.
These high level properties like liveness may be fairly obvious, but many of the attacks reviewed in this paper may fade into obscurity and become unknown to new community members.
We fear that new members may repeat these issues, and hope that a guideline for bridge construction may be established which would improve the quality of future bridges.

%% file: sections/conclusion.tex
\section{Conclusion}\label{sec:conclusion}

We explained several bridge attacks and suggested mitigations for most of them. Our work should not be seen as an exhaustive list of security issues to prevent, but rather as an insightful survey of principles that were exploited in the past, that the developers of bridges should keep in mind, and as an illustration of the complexity of the bridge systems, and the size of the attack surface that they expose for potential exploitation. 

%% file: sections/acknowledgement.tex
\paragraph*{Acknowledgement.}

The survey and research of bridge attacks was started at Quantstamp's Winter 2021 Research Retreat. The vulnerabilities, attacks, possible remedies, and solutions were discussed throughout the time with many security researchers and engineers at Quantstamp. The authors would like to thank especially their colleagues Kacper Bak, Sebastian Banescu, Mohsen Ahmadvand, and Marius Guggenmos for their insight and fruitful input that was incorporated in this paper.